# Beyond optical-field calculations: Experimental calibration of excitation density in microcavities

Yanmei He[1], Matheus Gomes Ferreira[1], Fan Wu[1, 2], Andrea Betti[3], Matteo Ghezzi[4], Edoardo Carraro[4], Tersilla Virgili[5], Dmitry Baranov[1], Giulio Cerullo[4,5], Mariangela Di Donato[3,6], Tönu Pullerits[1,*]

[1] Division of Chemical Physics and NanoLund, Lund University, P.O. Box 124, 22100 Lund, Sweden

[2] Department of Physics, College of Sciences, Shanghai University, Shanghai 200444, China

[3] European Laboratory for Non-Linear Spectroscopy (LENS), Via Nello Carrara 1, 50019 Sesto Fiorentino, Italy

[4] Department of Physics, Politecnico di Milano, Piazza Leonardo da Vinci 32, 20133 Milano, Italy

[5] Istituto di Fotonica e Nanotecnologia (IFN)-CNR, Piazza Leonardo da Vinci 32, 20133 Milano, Italy

[6] CNR-INO, Via Nello Carrara 1, 50019 Sesto Fiorentino, Italy

*Corresponding author: Tönu Pullerits. Email: tonu.pullerits@chem.lu.se

**Abstract**

Quantitative studies of cavity-modified photophysics require accurate determination of excitation intensity at the active layer, yet this quantity is commonly inferred from external pump fluence or intracavity optical-field calculations. Here, we present an *in-situ* calibration approach using perovskite colloidal quantum dots (QDs) as an intrinsic excitation-density probe. Fluence-dependent ultrafast pump-probe spectroscopy measurements on both perovskite QD films and microcavity-embedded films, combined with Poisson photon-absorption statistics and Auger-mediated multiexciton dynamics, enable the determination of the effective excitation intensity in a Fabry–Pérot optical microcavity. The cavity-to-film intensity scaling factor is 1.76 ± 0.10 near resonance and 1.26 ± 0.08 off resonance. Under near-resonant excitation, the experimental value is substantially lower than the standing-wave and transmission transfer-matrix predictions of 3.58 and 3.53. These results distinguish the effective excitation at the active layer from ideal optical-field enhancement, underscoring the need to account for sample-specific losses, structural inhomogeneity, and realistic measurement conditions. Our work provides an experimental framework for quantitative studies of cavity-modified excited-state dynamics.

## Introduction

Resonant optical microcavities provide a versatile platform for tailoring light–matter interactions by confining electromagnetic fields within small mode volumes.[1, 2, 3, 4, 5, 6] In the strong coupling regime, cavity photons and excitons hybridize to form light-matter states, known as polaritons.[1, 2, 3, 7] Beyond their fundamental importance in cavity quantum electrodynamics and polariton physics, these states have attracted increasing attention in chemistry and materials science because they can substantially reshape photophysical properties and ultrafast excited-state dynamics without altering molecular or crystal structure. Recent studies have shown that optical microcavities can influence excitation-energy transfer, nonlinear exciton loss channels such as exciton–exciton annihilation, and excited-state or charge-carrier lifetimes, thereby opening new ways to control photoinduced processes.[8, 9, 10, 11, 12, 13, 14, 15, 16] These findings have stimulated broad interest in microcavity-enabled control of photophysical and photochemical processes, including polariton condensation, polariton light-emitting diodes, lasing, and excitation-density-dependent phenomena.[3, 17, 18, 19, 20, 21, 22, 23, 24, 25, 26] A quantitative understanding of these processes, however, critically depends on the accurate determination of the effective excitation intensity at the active layer inside the optical microcavity (Fig. 1a).

Until now, determining the excitation intensity inside the microcavity has not been straightforward. In most experiments, the external incident pump intensity is used as the practical control parameter, whereas the effective excitation intensity at the active layer must be inferred indirectly. Existing descriptions rely predominantly on classical Fabry–Pérot optics to calculate internal field intensity in a lossy medium.[27] This framework is well established and remains indispensable for considering the counter-propagating standing wave in the planar cavity structure. Under realistic conditions, however, the intracavity excitation intensity can be influenced by many factors, including optical losses, finite field penetration into the mirrors, boundary conditions, mode mismatch, and spectral detuning between the pump and the cavity resonance. In microcavities containing an absorbing layer, these effects can modify the optical field distribution, particularly near the resonant position.[28] As a result, the intracavity excitation intensity is often difficult to determine directly, which complicates quantitative comparison of intra- and extracavity excited-state dynamics and obscures the interpretation of excitation-density-dependent phenomena. It is therefore of great importance to establish an experimental approach for *in-situ* measurement of the effective excitation intensity inside an optical microcavity, enabling quantitative analysis of cavity-modified excited-state dynamics across different material platforms.

Colloidal semiconductor quantum dots (QDs) provide an attractive model system for addressing this challenge because their multiexciton dynamics offer an intrinsic optical metric of excitation density. [29, 30, 31, 32, 33, 34, 35] When two or more excitons are generated in a single QD, efficient Auger recombination gives rise to a fast decay component (Fig. 1). The amplitude of such a component depends sensitively on excitation intensity, enabling precise determination of the absorption cross-section and average exciton occupancy, thereby connecting the ultrafast spectroscopic signals to the excitation density of QDs.[36, 37, 38, 39, 40] Here, we prepare $CsPbBr_3$ perovskite QD thin films and the Fabry-Pérot optical microcavities incorporating such films to establish an *in situ* calibration method of effective intracavity excitation intensity. Using intensity-dependent ultrafast pump-probe spectroscopy, the excited-state dynamics of bare and cavity-embedded films are systematically investigated. By analyzing the ground state bleaching (GSB) signals together with the Auger-recombination-related decay component as a function of pump fluence, we experimentally determine the excitation intensity in the optical microcavity under both near-resonant and off-resonant excitation. This calibration principle is more generally applicable to cavity systems in which the excitation density of the active material can be independently quantified through an excitation-dependent population response. Under this condition, the approach provides a transferable framework for quantitative comparison of intra- and extracavity excited-state dynamics under calibrated excitation conditions.

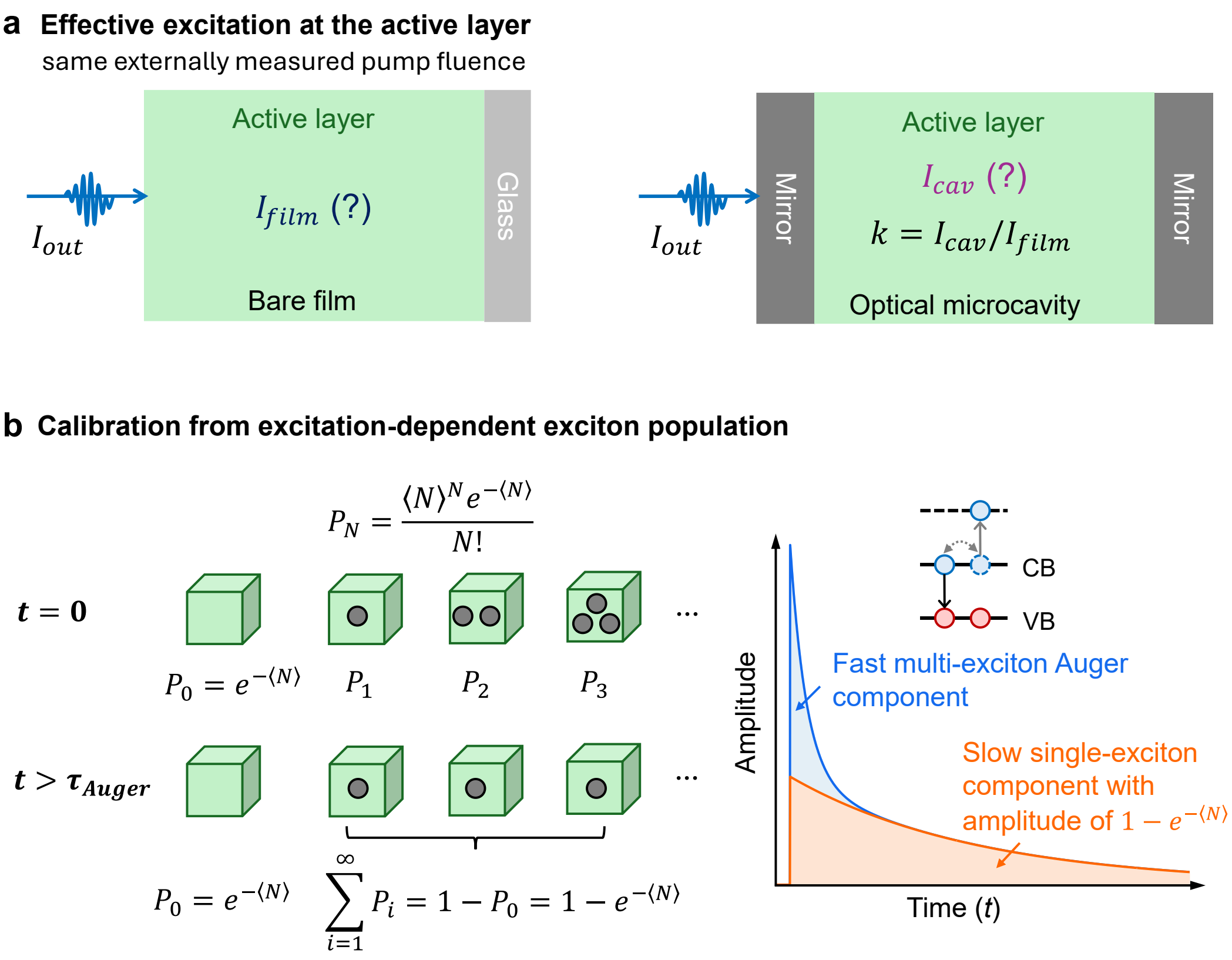

**Fig. 1 Concept of *in-situ* calibration of effective excitation intensity in an optical microcavity. a** Under the same external pump fluence, the effective excitation intensities at the active layer in the bare film ($I_{film}$) and optical microcavity ($I_{cav}$) can be different. Their relative excitation-intensity scaling factor is defined as $k = I_{cav} / I_{film}$. **b** Following pulsed excitation, the exciton occupation per QD is assumed to follow a Poisson distribution with a mean value of $\langle N\rangle$. Multiexciton populations undergo rapid Auger recombination, leaving a slow single-exciton contribution with the amplitude of $1 - \mathrm{e}^{-\langle N\rangle}$. Decomposition of ultrafast dynamics into fast multiexciton and slow single-exciton components enables the determination of $\langle N\rangle$ and, consequently, the effective excitation intensity inside the optical microcavity.

## Results

### Microcavity structure and optical characteristics

The $\lambda$-thick Fabry–Pérot microcavity studied here is schematically illustrated in Fig. 2a. It consists of a spin-coated $CsPbBr_3$ QDs (6.3 nm, Fig. S1)/polymethyl methacrylate (PMMA) blend film embedded between a pair of 30 nm Ag mirrors with 15 nm $SiO_2$ spacer layers. We first measure angle-resolved transmission spectra of the microcavity sample. As shown in Fig. 2b and Fig. S2, two dispersive branches are clearly resolved and can be assigned to the upper polariton (UP) and lower polariton (LP) branches, indicating strong coupling. The measured dispersion is well fitted by a two-coupled-oscillator model, confirming hybridization between the cavity photon mode and the QD excitonic transition. A pronounced anti-crossing is observed at an incidence angle of ~25 degrees, with the extracted Rabi splitting ($\hbar\Omega$) energy of ~124 meV. Consistent with this interpretation, the Hopfield coefficients derived from the model show the largest photon–exciton admixture near this angle (Fig. S3). The comparison of the QD film absorption and microcavity transmission spectra further corroborate the spectral overlap underlying the polariton formation (Fig. 2c).

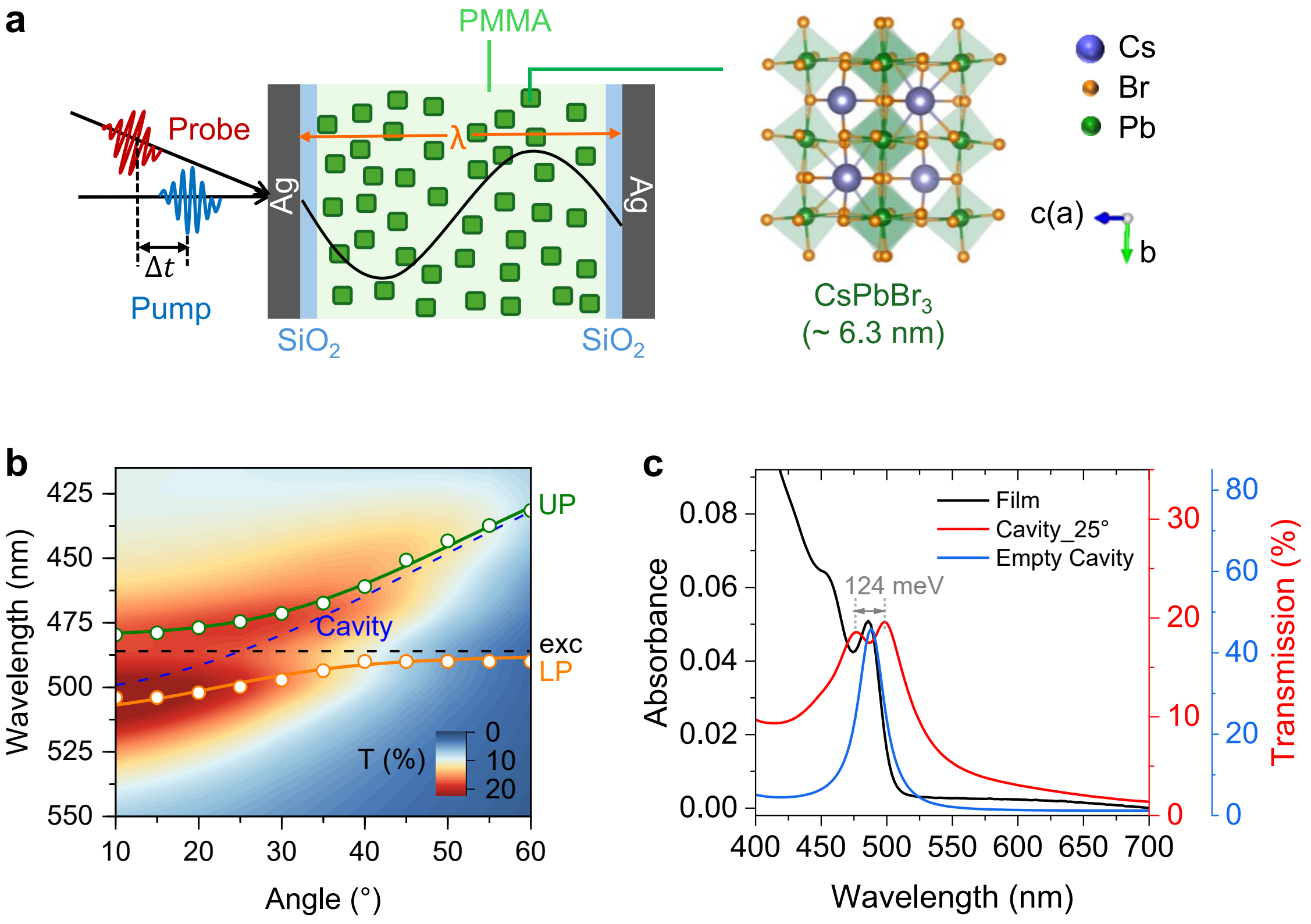

**Fig. 2 Strong exciton-photon coupling of $CsPbBr_3$ quantum dots in a Fabry-Pérot microcavity. a** Schematic of a semitransparent $\lambda$ microcavity comprising a pair of $SiO_2$ (15 nm) /Ag (30 nm) mirrors and a thick PMMA layer (220 nm) containing $CsPbBr_3$ QDs (~6.3 nm) as the active layer. The pump-probe geometry is shown with the nominal incident angle of photoexcitation. **b** Experimentally measured angle-resolved transmission map overlaid with the fitting results based on a coupled-oscillator model. The dots mark the positions of the polariton branches, while the exciton and photon energies are indicated by black and grey dashed lines, respectively. **c** Steady-state absorption spectrum of the bare QD film and transmission spectrum of the microcavities with and without QDs.

**Excitation-intensity calibration near cavity resonance**

To quantify and compare the excitation conditions in a cavity and bare film, we perform fluence-dependent ultrafast pump-probe experiments in both cases using a tunable ultrashort pump pulse for excitation and a broadband white-light supercontinuum for probing. We first set the pump wavelength to 490 nm, close to the excitonic absorption peak of the QD film, with the pump beam oriented perpendicular to the sample surface. As shown in Fig. 3a-b and Fig. S4, the bare film and microcavity exhibit broadly similar derivative-like spectral features around the excitonic transition, while the cavity spectrum shows a larger signal amplitude and an additional weak positive feature on the red side of the bleach band. The latter may partly arise from cavity-induced spectral reshaping, as photoinduced changes in the optical response of the QD layer can modify the transient transmission response of the cavity.[41, 42] A definitive assignment of this feature, however, requires further investigation. As described in Fig. 1, upon pulsed excitation, photon absorption by individual QDs is stochastic, giving rise to a Poisson distribution of exciton occupancies across the QD ensemble. QDs containing only one exciton follow slow single-exciton relaxation, whereas QDs containing two or more excitons initially undergo rapid Auger multiexciton decay, in which one electron-hole pair recombines non-radiatively and transfers its energy to another carrier within the same QD. This highly excited hot carrier rapidly relaxes to the band edge. Auger decay in QDs is significantly faster than single exciton decay. This means that after Auger decay is completed, all initially excited QDs (multiply or singly) have just one remaining excitation. The fluence-dependent GSB kinetics reflect the changed population balance between the slow single-exciton decay and the fast Auger-mediated multiexciton decay. By analysing the GSB kinetics, we monitor the time-dependent Pauli blocking of the excitonic transition and analyze the fluence-dependent recovery using a population-based rate-equation model that incorporates Poisson-distributed exciton generation and Auger-mediated multiexciton depopulation, allowing us to precisely determine the average number of photogenerated excitons per QD (Supplementary Note 1).[30, 36, 37] As shown in Fig. 3c-d, at low excitation fluence, the GSB kinetics of both samples show a nearly single-exponential decay. As the pump fluence increases, hundreds of picoseconds rapid decay components are observed at early times which become more pronounced at higher pump fluence, demonstrating the occurrence of Auger recombination. At matched average exciton occupancies, the cavity sample exhibits a shorter fast multiexciton-associated decay than the bare film (Fig. S5-6, Supplementary Note 2). Importantly, the exciton-occupancy analysis relies on the fluence dependence of the component amplitudes, particularly that of the longest-lived component, rather than on the decay time of the faster multiexciton-associated component. Therefore, the observed change in the fast decay dynamics does not influence the occupancy determination provided multiexciton relaxation is complete in the time window used to extract the slow component. Establishing the microscopic origin of the effect requires further investigation.

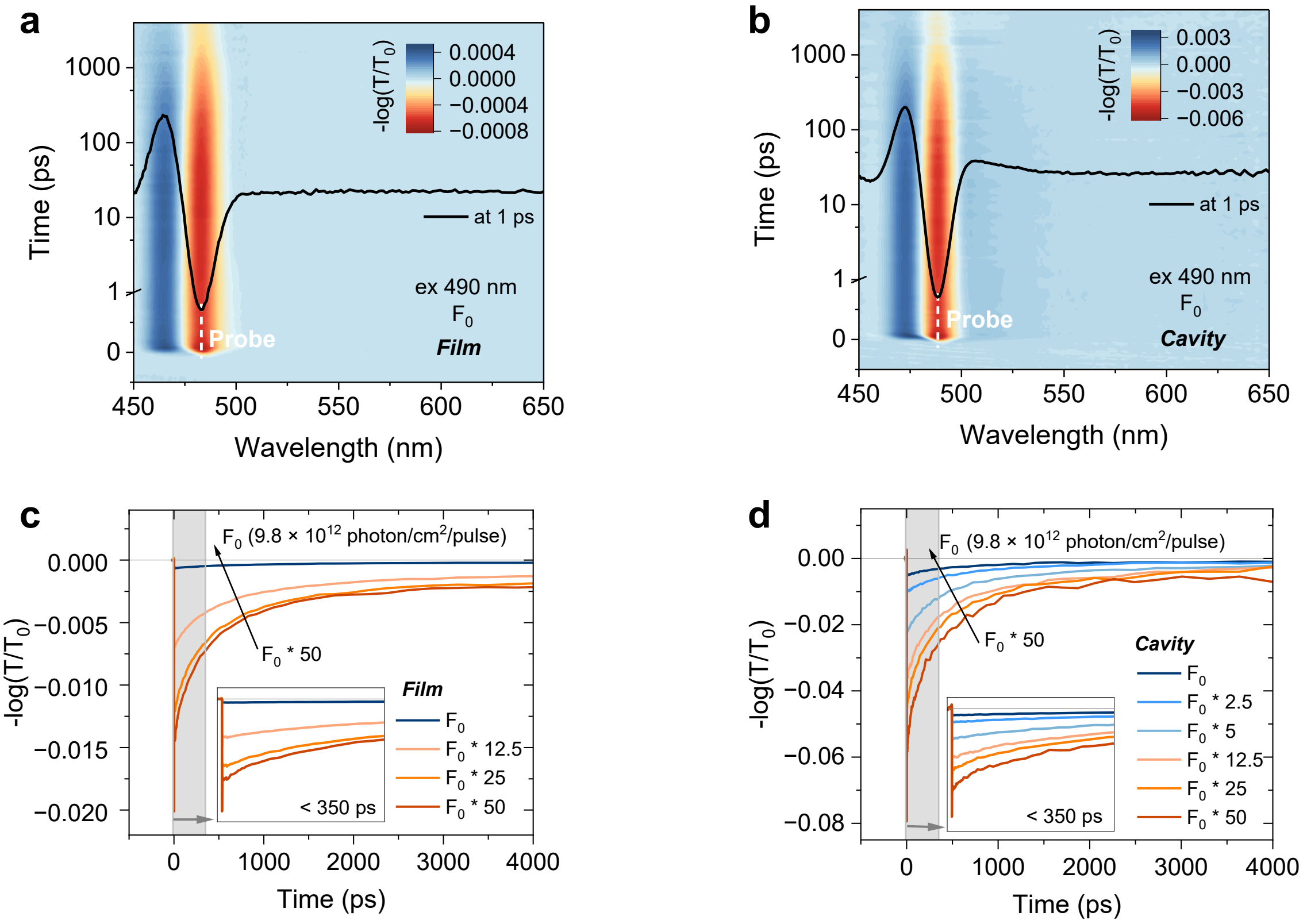

**Fig. 3 Fluence-dependent ultrafast pump-probe measurements of QD film and microcavity sample.** Pseudocolor representation of broadband pump-probe spectra of **a** bare film and **b** microcavity sample excited at 490 nm at a nominal incident angle. The inset figure is the spectrum at 1 ps. Fluence-dependent kinetics at the GSB feature of **c** bare film and **d** microcavity sample. The first 350 ps of kinetics are shown to present the fast Auger recombination.

To quantify the excitation density at the QD layer, we now estimate the average exciton occupancy per QD from the fluence-dependent GSB kinetics using the mentioned Poisson model. The GSB signal amplitude is analyzed as a function of pump fluence to determine the effective absorption cross-section of the bare QD film, yielding a value of $9.14 \times 10^{-15}$ $cm^2$ (Fig. 4a, Supplementary note 1).[39] Fig. 4b shows the corresponding Poisson-distributed probability of generating $N$ excitons per QD in the bare film at different excitation fluences, with the distribution progressively broadening and shifting toward higher exciton numbers as the average exciton occupancy ($\langle N \rangle$) increases. We then apply the same analysis to the corresponding microcavity sample, assuming that the intrinsic absorption cross-section of the QDs remains unchanged upon incorporation into the cavity. This allows the effective excitation intensity at the QD layer within the cavity to be expressed as a scaled form of the intensity in the film, with the scaling factor extracted directly from the microcavity sample dataset. The resulting scaling factor is ~ 1.76 ± 0.10, indicating that under near-resonant excitation, the effective excitation intensity at the QD layer in the cavity is approximately twice that in the corresponding bare film under the same incident pump fluence (Fig. 4c). Consistent with this result, Fig. 4d shows that the exciton-number distribution in the microcavity is shifted to substantially higher $\langle N \rangle$ values than in the bare film under the same nominal excitation conditions. This calibration will enable a rigorous quantitative assessment of how the microcavity modifies the excited-state dynamics relative to the bare film.

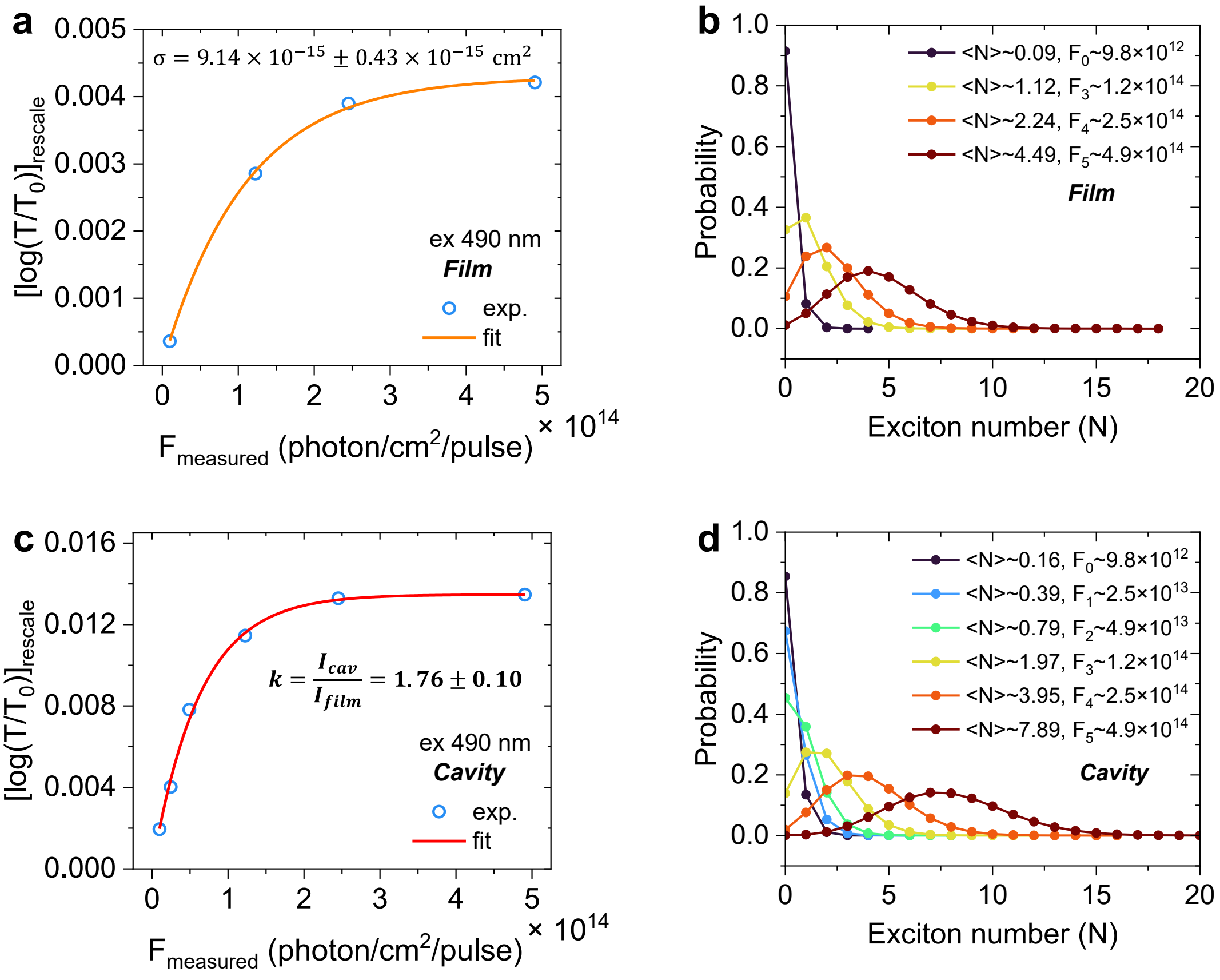

**Fig. 4 Effective absorption cross-section and cavity-to-film intensity scaling factor at 490 nm.** Bleaching signal amplitude rescaled (see Supplementary note 1 for details) from $\Delta t > 1$ ns as a function of excitation fluence (F, photon/cm$^2$/pulse) for **a** bare film and **c** microcavity

sample excited at 490 nm. The inset shows the effective absorption cross-section ($\sigma$) and scaling factor ($k$) of excitation intensity inside the cavity ($I_{cav}$) and bare film samples ($I_{film}$). Poisson distribution of multiple exciton population after excitation: **b** bare film; **d** microcavity sample.

**Excitation-intensity calibration off cavity resonance**

In planar Fabry–Pérot microcavities, light–matter coupling is maximized when the active material is positioned at the antinodes of the cavity electric field.[43, 44, 45] Under optical excitation, the intracavity field intensity, and therefore the effective photon fluence, is expected to be highly sensitive to the spectral detuning between the pump wavelength and the cavity resonance. We carry out an additional set of fluence-dependent ultrafast pump-probe measurements using 470 nm excitation, which is blue-detuned from the QD excitonic transition. The pump-probe spectra of the bare film and microcavity sample are qualitatively similar to those obtained under 490 nm excitation (Fig. S7-8). The threshold for the fast Auger recombination in the microcavity sample shifts toward slightly lower incident pump fluences compared to that of the bare film (Fig. 5). This observation is further supported by a higher average exciton occupancy per QD in the microcavity, primarily arising from the internal excitation intensity enhancement (Fig. S9). Using the same Poisson-based statistical framework described above, we first fit the intensity-dependent bleaching amplitudes of the bare film to extract the corresponding QD effective absorption cross-section. This value is then used to fit the microcavity data, thereby determining the intracavity excitation intensity scaling factor. The extracted scaling factor is ~1.26 ± 0.08, substantially smaller than that obtained under 490 nm excitation, indicating that the effective excitation intensity in the QD layer is reduced as the pump is tuned further away from resonance. Taken together, these comparative measurements at 470 and 490 nm excitation show that the apparent intensity threshold for Auger recombination is primarily governed by the cavity-modulated effective excitation intensity.

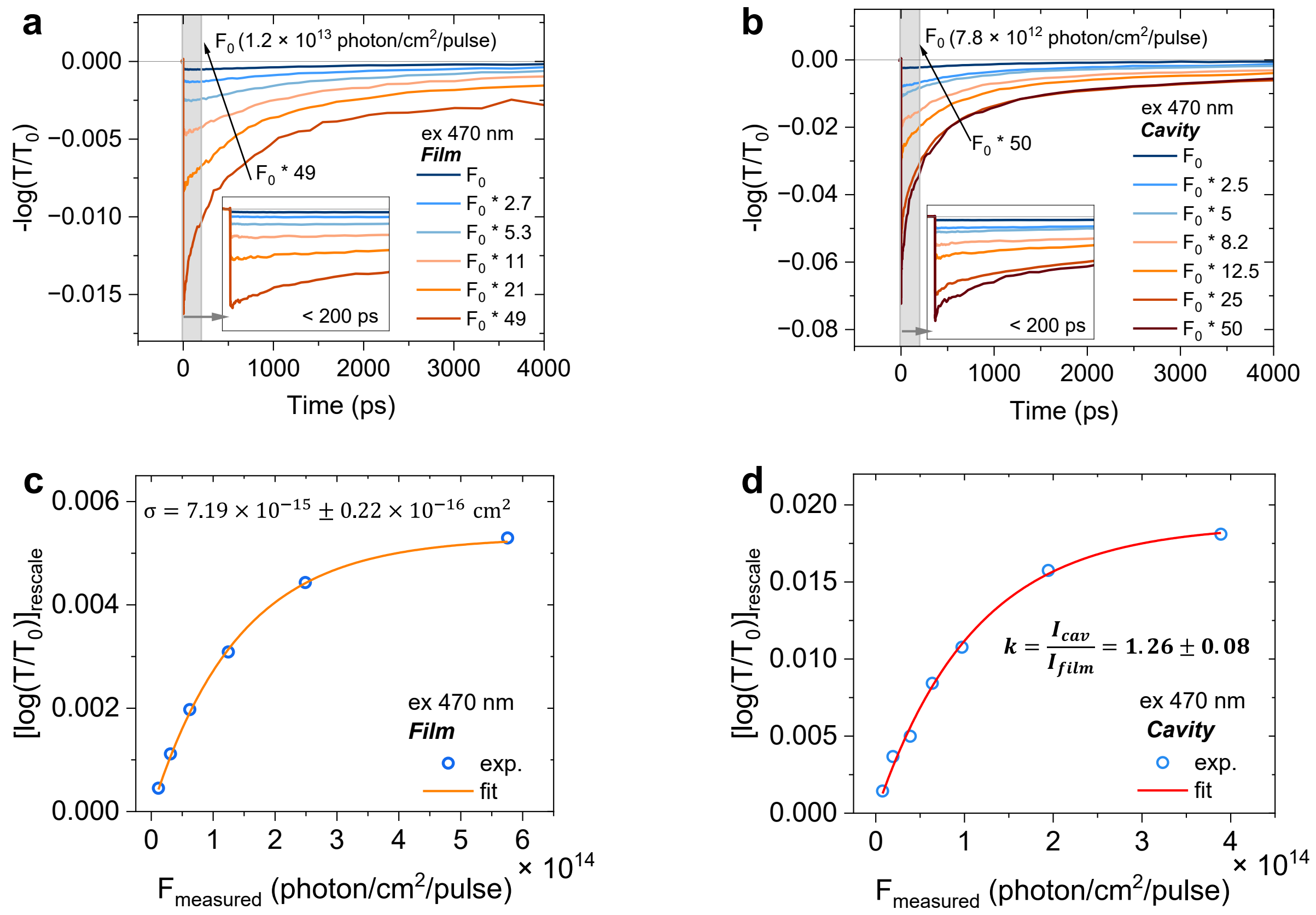

**Fig. 5 Effective absorption cross-section and cavity-to-film excitation calibration at 470 nm.** Intensity-dependent kinetics at the GSB feature of **a** bare film and **b** microcavity sample. The first 200 ps of kinetics are shown to present the fast Auger recombination. Bleaching signal amplitude rescaled from $\Delta t$ > 1 ns as a function of excitation fluence for **c** bare film and **d** microcavity sample pumped at 470 nm. The inset shows the effective absorption cross-section ($\sigma$) and scaling factor ($k$) of excitation intensity inside the cavity ($I_{cav}$) and bare film samples ($I_{film}$).

## Comparison with optical-field models

To determine how the experimentally extracted intensity scaling factor relates to conventional optical models, we compare the following three methods (Fig. 6): *in-situ* ultrafast pump-probe measurements, an ideal counter-propagating standing-wave model based on the cavity energy relations described by Nagourney,[27] and transfer-matrix method (TMM) calculations.[46] These approaches quantify the excitation enhancement inside the microcavity through different observables and assumptions. The standing-wave model gives a simplified analytical description based on the coherent superposition of counter-propagating fields. The TMM calculation quantifies the coherent optical response of a planar multilayer stack using the measured optical constants and layer thicknesses. By contrast, the pump–probe approach derives the scaling factor from the fluence–dependent Auger dynamics, yielding an experimental calibration of the QD-layer effective excitation intensity. Details of these methods are provided in Supplementary Notes 3–4 and Figs. S10–S15.

Under the near-resonant conditions ($\lambda_p$ = 490 nm), we have extracted an experimental scaling factor of 1.76 ± 0.10. In contrast, the ideal counter-propagating standing-wave model predicts intensity enhancements of 3.58. The idealized counter-propagating standing-wave model gives the highest value compared to experimental results, because it represents an idealized interference limit and does not explicitly resolve the dissipative multilayer structure, including layer-dependent absorption, mirror phase shifts, field penetration, multiple internal reflections, and thickness-dependent phase accumulation. In a complementary TMM analysis, the microcavity active-layer field factor was normalized by the corresponding value calculated for a bare film on glass, yielding relative cavity-induced intensity enhancements of 3.53 at 490 nm (Fig. S14-15). The TMM calculation incorporates these multilayer optical effects through the experimentally constrained optical constants and layer thicknesses, which could account for the lower value relative to the standing-wave model. Nevertheless, it remains a linear coherent optical description of an ideal planar multilayer structure and does not fully capture structural inhomogeneity, interfacial roughness, and scattering. The value calculated from the ultrafast pump-probe measurement, therefore, represents an effective experimental scaling factor for the intracavity excitation intensity under the present measurement conditions.

A different trend is observed under off-resonant excitation ($\lambda_p$ = 470 nm), where the pump wavelength is further detuned from cavity resonance. The experimental scaling factor decreases to 1.26 ± 0.08, consistent with the weaker field buildup inferred from the intensity-dependent kinetics. This value is lower than the value calculated from the counter-propagating standing-wave model (1.65). This discrepancy at high detuning should be interpreted as a theoretical limitation of the simplified standing-wave approximation, which is consistent with its substantial overestimation under near-resonant excitation. The TMM calculation yields a smaller intensity scaling factor of 1.14. This lower value can be rationalized by the stronger detuning from the cavity mode and the wavelength-dependent phase condition of the planar multilayer structure, which together reduce the calculated pump-intensity enhancement at the QD layer under off-resonant excitation. The discrepancy does not necessarily indicate an incorrect optical calculation but rather reflects the distinction between a linear intensity-scaling factor calculated for an ideal planar multilayer cavity and an experimentally calibrated effective excitation-intensity scaling factor derived from Auger dynamics. These comparisons suggest that excitation conditions in dissipative microcavities cannot always be inferred quantitatively from optical calculations alone. The standing-wave and TMM models describe the expected intracavity intensity enhancement within their respective optical assumptions, whereas the *in-situ* pump–probe method calibrates the effective excitation intensity at the active layer through the observed fluence-dependent Auger-related kinetics. For the quantitative analysis of cavity-modified excited-state dynamics, the experimentally derived scaling factor therefore provides an empirical calibration of the effective excitation intensity at the active layer.

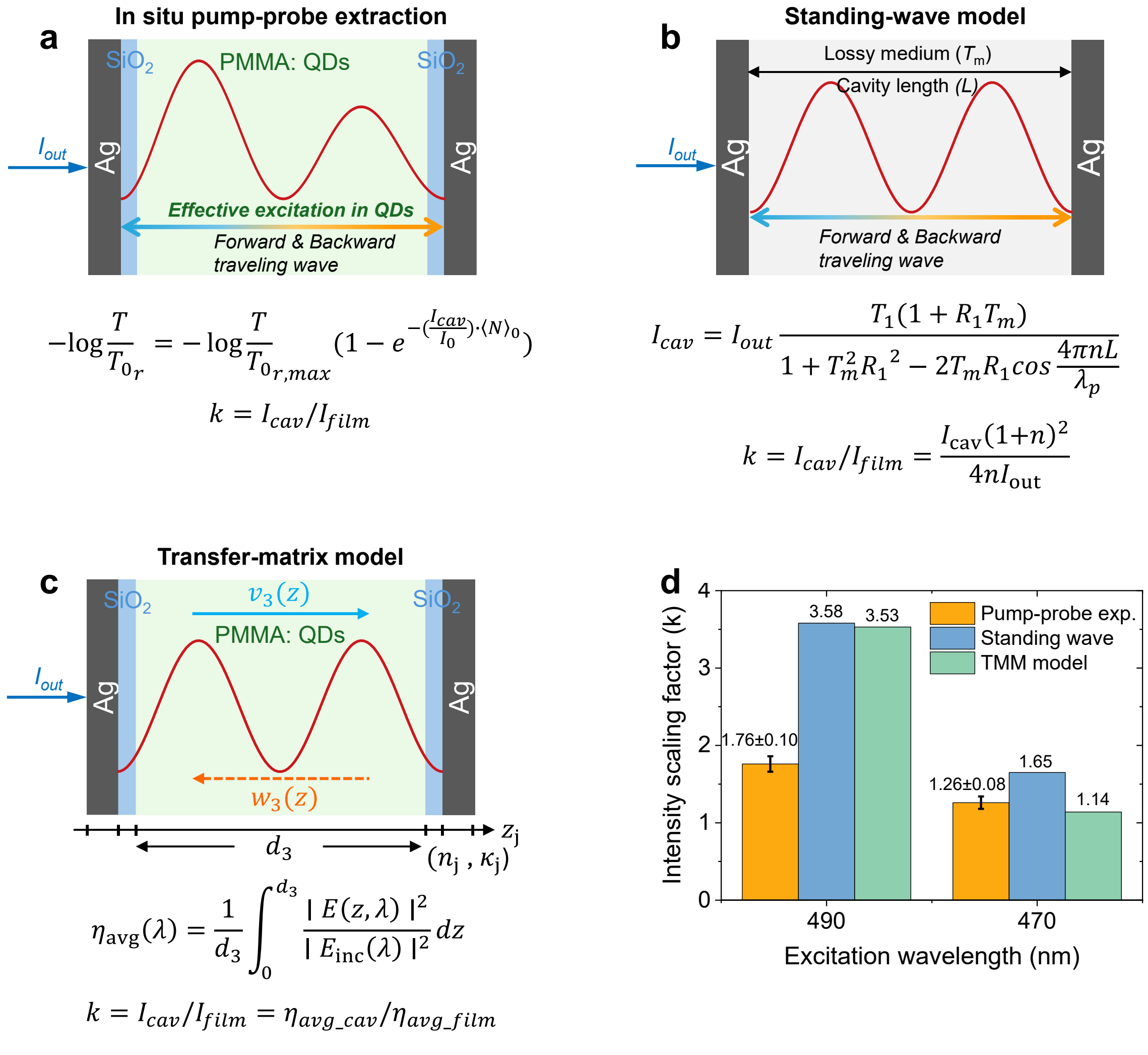

**Fig. 6 Comparison of experimental intracavity-intensity calibration with optical-field models.** The red sine wave represents the intensity distribution across the cavity length. $I_{out}$ is the external incident excitation intensity. **a** Experimental in-situ calibration of the effective excitation intensity using ultrafast pump-probe spectroscopy and photon-absorption statistics. $I_0$ is the lowest external excitation intensity. **b** Ideal standing-wave model formed by counter-propagating waves between two mirrors. $n$, $T_1$, $T_{\text{med}}$, and $R_1$ denote the refractive index, mirror

transmittance, medium transmittance, and mirror reflectance amplitudes, respectively. **c** TMM calculation of the active-layer field profile from the coherent superposition of forward- and backward-propagating waves ($v_3$, $w_3$). $d$, $E$, $\eta_{\mathrm{avg}}$ and $k$ represent layer thickness, electric field, averaged electric-field factor, and intensity scaling factor, respectively. **d** Cavity-to-film intensity scaling factor calculated from theoretical and experimental methods.

## Discussion

By using the fluence-dependent kinetics of perovskite QDs as an intrinsic probe for exciton population, we establish an experimental calibration method of the effective excitation intensity at the QD layer. Within this scheme, the effective absorption cross-section is first determined from the bare-film bleaching signal using the Poisson model for photon absorption and is then used to extract the effective excitation intensity in the corresponding microcavity sample. This calibration enables quantitative comparison between bare film and cavity sample under matched effective excitation densities. The effective intracavity intensity depends strongly on pump detuning. Under near-resonant excitation, the cavity produces a substantially larger enhancement of the excitation intensity at the active layer than under more off-resonant pumping, which accounts for the observed shift in the external intensity required to reach the multiexciton recombination regime. The comparison between 490 and 470 nm excitation, therefore, shows that the apparent threshold for Auger-dominated dynamics is governed primarily by the effective intracavity excitation intensity.

Comparison with optical-field models further clarifies the distinction between calculated coherent-field enhancement and experimentally calibrated effective excitation. The ideal counter-propagating standing wave model acts as a simplified optical reference model compared to the experimentally extracted excitation intensity for both near-resonant and detuned excitation regimes examined here. The discrepancy becomes increasingly pronounced near resonance, since the sample-specific optical losses, including reflection phase shifts and field penetration into the metallic mirrors, produce the asymmetric damping of the intracavity field. Although the TMM calculation provides a more detailed optical description of the experimentally constrained multilayer structure, it remains a coherent optical-field description that ignores the structural inhomogeneity and real measurement conditions. It is not equivalent to the experimentally calibrated excitation scaling factor obtained from Auger dynamics. This work therefore highlights the need for experimental calibration of the effective excitation intensity at the active layer in quantitative studies of dissipative microcavities. More generally, this approach provides a practical route to disentangle cavity-induced changes in excitation conditions from intrinsic modifications of excited-state dynamics due to intracavity intensity enhancement. Although demonstrated here using multiexciton statistics and Auger recombination in perovskite quantum dots, the underlying calibration concept can, in principle, be extended to other cavity systems in which the excitation density can be independently quantified from a well-defined excitation-dependent spectroscopic signal.

## Methods

### Synthesis and isolation of ~6.3 nm $CsPbBr_3$ quantum dots (QDs)

The synthesis and isolation of the ~6.3 nm $CsPbBr_3$ QDs followed a modified protocol from Gomes Ferreira et al.[47, 48] Initially, a cesium oleate stock solution was prepared by dissolving 400 mg cesium carbonate ($Cs_2CO_3$) in 1.75 mL of oleic acid (OA) and 15 mL of 1-octadecene (ODE) at 150 ºC under continuous stirring until the mixture became transparent.

For the QDs synthesis, 75 mg of lead (II) bromide ($PbBr_2$) and 180 mg of zinc bromide ($ZnBr_2$) were combined in a 20 mL glass vial with 5 mL ODE, 2 mL OA, and 2 mL oleylamine. The mixture was heated in an aluminum block to ~130ºC until the salts were completely solubilized, appearing as a translucent yellow solution. The temperature was then raised to 165 ºC, followed by a rapid injection of 0.4 mL cesium oleate stock. The reaction was immediately quenched by removing the vial from the heat and cooling it to ~25 ºC under stirring.

To isolate the QDs, the crude mixture was equally divided into three tubes and then centrifuged at 3500 rpm for 10 minutes. The supernatant in each tube was collected and mixed with 9 mL of ethyl acetate to induce precipitation, followed by further centrifugation at 3500 rpm for 2 minutes. After discarding the supernatant and removing residual liquid via a brief 1-min centrifugation, the resulting precipitates from the three tubes were redispersed in 100 µL of anhydrous toluene in total. Here we prepared two batches, resulting in a final volume of 200 µL for the film and cavity preparation.

### Fabry-Pérot cavity fabrication

The optical cavities were fabricated on glass substrates (15 × 15 $mm^2$). Before deposition, the substrates were sequentially cleaned by sonication in alkaline solution (0.5% Hellma-nex in deionized water, 10 min), deionized water (10 min, twice), acetone (10 min), and isopropanol (10 min), and then treated with oxygen plasma for 1 min. A semi-transparent Ag layer (30 nm) and a $SiO_2$ spacer layer (15 nm) were subsequently deposited by vacuum sputtering (AJA Orion 5). To form the active layer, PMMA was dissolved in anhydrous toluene at 80 mg $mL^{-1}$ and stirred overnight at 60 °C. The QD solution was then blended with the PMMA solution at a volume ratio of 2:1 (40 μL: 20 μL), and the mixture was spin-coated onto the $SiO_2$/Ag-coated substrate at 3000 rpm for 30 s using a Laurell Technologies WS-650 spin coater. A second $SiO_2$/Ag mirror pair with identical thicknesses was deposited on top of the active layer to complete the cavity structure. The as-prepared cavity was stored in a glove box before experiments to prevent oxidation and sample degradation. As a control, a QD film was prepared by spin-coating the same precursor mixture onto a cleaned glass substrate under identical conditions, but without the top and bottom mirrors and $SiO_2$ layers.

**Morphology characterization and steady-state spectroscopy**

Bright-field transmission electron microscopy (TEM) was used to characterize the $CsPbBr_3$ QDs. Samples were prepared by drop-casting 10 μL of the concentrated QD dispersion onto a lacey carbon support film on a 300-mesh copper grid, followed by drying. Imaging was performed using a JEOL JEM-3000F high-resolution transmission electron microscope operated at 300 kV. Steady-state spectra were recorded on a standard spectrophotometer (Lambda 1050, PerkinElmer) fitted with appropriate accessories. The angle-resolved transmission measurements were carried out using a variable-angle accessory. The film thickness was measured using a Bruker Dektak XT Stylus Profilometer.

**Ultrafast pump-probe measurements**

Ultrafast pump-probe measurements were performed on a home-built spectroscopic setup. A Solstice regenerative amplifier (Spectra-Physics; 8 W, 796 nm, 60 fs, 4 kHz), seeded by a Mai Tai SP femtosecond oscillator (Spectra-Physics), served as the laser source. The output beam was split into pump and probe components with power ratios of 95% and 5%, respectively. The probe continuum was generated by focusing 1350 nm pulses from a collinear optical parametric amplifier (TOPAS-C, Light Conversion) onto a thin $CaF_2$ plate, yielding a white-light spectrum in the 430-1000 nm range. This probe beam was divided into a signal beam passing through the sample and a reference beam directed to the detector. The pump pulse (80 mW, 470 or 490 nm, 100 fs) was generated by a second TOPAS-C system (Light Conversion) through sum-frequency generation of the 800 and 1300 nm beams.

To suppress pump-scattering artifacts, the pump and probe polarizations were set orthogonal to each other. The excitation fluence was adjusted sequentially using neutral-density filters. The pump beam had a spot radius of approximately 200 μm at the sample position. To minimize photodamage and cumulative photocharging during the measurements, the sample was spatially translated by 500 μm either after completion of each time-delay scan or at every time-delay step, depending on the acquisition protocol. The translation distance was larger than the pump-spot diameter (~400 μm), thereby minimizing repeated excitation of the same sample area. In the scan mode, the same number of laser shots was averaged at each scan for all excitation fluences. In the point mode, the number of laser shots was adjusted according to the excitation fluence, with fewer shots used at higher fluences to further reduce the cumulative excitation dose at each sample position.

**Coupled oscillator model**

To describe the polariton dispersion observed in the QD-embedded microcavity, the energy branches were analyzed using a 2-by-2 coupled-oscillator Hamiltonian:

$$\begin{pmatrix} E_C(\theta) & V \\ V & E_{exc} \end{pmatrix} \begin{pmatrix} \alpha \\ \beta \end{pmatrix} = E \begin{pmatrix} \alpha \\ \beta \end{pmatrix}$$

Where $E$ represents the eigenvalues corresponding to the upper and lower polariton branches (UP and LP). The terms $E_C(\theta)$ and $E_{exc}$ denote the energies of the cavity mode and the nanocrystal exciton, respectively, while $V$ represents the coupling strength. The mixing coefficients, α and β, define the eigenvectors of the strongly coupled system, with the corresponding Hopfield coefficients ($|\alpha|^2$ and $|\beta|^2$) quantifying the photon and exciton fractions within each polariton state. The dispersive nature of the cavity mode $E_C(\theta)$ is governed by the incident angle:

$$E_C(\theta) = \frac{E_0}{\sqrt{1-(\frac{sin\theta}{n_{eff}})^2}}.$$

Here $E_0$ is the cavity energy at $\theta$ = 0°. $n_{eff}$ is the effective refractive index. The coupling strength $V$ is related to the Rabi splitting $\hbar\Omega$, when $E_C(\theta) = E_{exc}$, it is given by $V = \hbar\Omega/2$. Here, $V$ and $n_{eff}$ are obtained by fitting the experimental dispersion curves to the Hamiltonian eigenstates using the Trust-Region-Reflective algorithm.

**Data availability**

The data supporting the findings of this study are available from the corresponding author upon reasonable request.

**Acknowledgements**

We acknowledge financial support from the Swedish Foundation of Strategic Research (IS24-0005), Swedish Energy Agency (50709-1), Olle Engkvists Foundation (235-0422), Light and Materials profile area at Lund University (Young Investigator Synergy Award, 2025), the Royal Physiographic Society of Lund, and NanoLund. The work of M.G.F. and D.B. was funded by the European Union (ERC Starting Grant PROMETHEUS, project no. 101039683). Views and opinions expressed are, however, those of the authors only and do not necessarily reflect those of the European Union or the European Research Council Executive Agency. Funded by the European Union under HE-GA 101131771 Lasers4EU [WP4]. Views and opinions expressed are, however, those of the author(s) only and do not necessarily reflect those of the European Union. The European Union cannot be held responsible for them.

**Author contribution**

T.P., M.D.D., and G.C. coordinated the project. Y.H. fabricated cavity/film samples and performed steady-state spectroscopy, angle-resolved and ultrafast pump-probe measurements. M.G.F. prepared the perovskite quantum dots with the supervision of D.B.. F.W. contributed to the discussion on microcavity preparation and angle-resolved measurements. A.B., M.G., E.C., T.V., M.D.D., and G.C. contributed to the discussions on the ultrafast pump-probe measurements. T.P. and Y.H. analyzed the data and formulated the conclusions. Y.H. prepared the first draft of the manuscript. All authors contributed to the discussion of the data and revision of the manuscript.

**Competing interests**

The authors declare no competing interests.

**References**

1. Hirai, K., Hutchison, J. A. & Uji-i, H. Molecular chemistry in cavity strong coupling. *Chem. Rev.* **123,** 8099-8126 (2023).
2. Xiang, B. & Xiong, W. Molecular polaritons for chemistry, photonics and quantum technologies. *Chem. Rev.* **124,** 2512-2552 (2024).
3. Garcia-Vidal, F. J., Ciuti, C. & Ebbesen, T. W. Manipulating matter by strong coupling to vacuum fields. *Science* **373,** eabd0336 (2021).
4. Ebbesen, T. W. Hybrid light–matter states in a molecular and material science perspective. *Acc. Chem. Res.* **49,** 2403-2412 (2016).

5. Xiong, W. Molecular vibrational polariton dynamics: What can polaritons do? *Acc. Chem. Res.* **56,** 776-786 (2023).
6. Simpkins, B. S., Dunkelberger, A. D. & Vurgaftman, I. Control, modulation, and analytical descriptions of vibrational strong coupling. *Chem. Rev.* **123,** 5020-5048 (2023).
7. Lidzey, D. G. *et al.* Strong exciton–photon coupling in an organic semiconductor microcavity. *Nature* **395,** 53-55 (1998).
8. Wu, F. *et al.* Optical cavity-mediated exciton dynamics in photosynthetic light harvesting 2 complexes. *Nat. Commun.* **13,** 6864 (2022).
9. Wu, F., Nguyen- Phan, T. C., Cogdell, R. & Pullerits, T. Efficient cavity-mediated energy transfer between photosynthetic light harvesting complexes from strong to weak coupling regime. *Nat. Commun.* **16,** 5300 (2025).
10. Stranius, K., Hertzog, M. & Börjesson, K. Selective manipulation of electronically excited states through strong light–matter interactions. *Nat. Commun.* **9,** 2273 (2018).
11. Akselrod, G. M., Tischler, Y. R., Young, E. R., Nocera, D. G. & Bulovic, V. Exciton-exciton annihilation in organic polariton microcavities. *Phys. Rev. B* **82,** 113106 (2010).
12. Tibben, D. J. *et al.* Molecular energy transfer under the strong light–matter interaction regime. *Chem. Rev.* **123,** 8044-8068 (2023).
13. Coles, D. M. *et al.* Polariton-mediated energy transfer between organic dyes in a strongly coupled optical microcavity. *Nat. Mater.* **13,** 712-719 (2014).
14. Zhong, X. *et al.* Energy transfer between spatially separated entangled molecules. *Angew. Chem. Int. Ed.* **56,** 9034-9038 (2017).
15. Georgiou, K., Jayaprakash, R., Othonos, A. & Lidzey, D. G. Ultralong-range polariton-assisted energy transfer in organic microcavities. *Angew. Chem. Int. Ed.* **60,** 16661-16667 (2021).
16. Hou, S. *et al.* Ultralong-range energy transport in a disordered organic semiconductor at room temperature via coherent exciton-polariton propagation. *Adv. Mater.* **32,** 2002127 (2020).
17. Kéna-Cohen, S. & Forrest, S. R. Room-temperature polariton lasing in an organic single-crystal microcavity. *Nat. Photon.* **4,** 371-375 (2010).
18. Zhao, H., Arneson, C. E., Fan, D. & Forrest, S. R. Stable blue phosphorescent organic LEDs that use polariton-enhanced Purcell effects. *Nature* **626,** 300-305 (2024).

19. Liu, B., Michail, E., He, G., Sfeir, M. Y. & Forrest, S. R. Ultrafast dynamics of Bloch surface wave polaritons in large-area 2D semiconductor monolayers at room temperature. *Adv. Mater.* **36,** 2404286 (2024).
20. Georgakilas, I. *et al.* Room-temperature cavity exciton-polariton condensation in perovskite quantum dots. *Nat. Commun.* **16,** 5228 (2025).
21. Kędziora, M. *et al.* Predesigned perovskite crystal waveguides for room-temperature exciton–polariton condensation and edge lasing. *Nat. Mater.* **23,** 1515-1522 (2024).
22. Su, R. *et al.* Perovskite semiconductors for room-temperature exciton-polaritonics. *Nat. Mater.* **20,** 1315-1324 (2021).
23. Ohtani, K. *et al.* An electrically pumped phonon-polariton laser. *Sci. Adv.* **5,** eaau1632 (2019).
24. Wang, M., Hertzog, M. & Börjesson, K. Polariton-assisted excitation energy channeling in organic heterojunctions. *Nat. Commun.* **12,** 1874 (2021).
25. Dietrich, C. P. *et al.* An exciton-polariton laser based on biologically produced fluorescent protein. *Sci. Adv.* **2,** e1600666 (2016).
26. Polak, D. *et al.* Manipulating molecules with strong coupling: Harvesting triplet excitons in organic exciton microcavities. *Chem. Sci.* **11,** 343-354 (2020).
27. Nagourney, W. *Quantum electronics for atomic physics and telecommunication*. (OUP Oxford, 2014).
28. Spencer, M. S. *et al.* Electro-optic cavities for in-situ measurement of cavity fields. *Light Sci. Appl.* **14,** 69 (2025).
29. Klimov, V. I., Mikhailovsky, A. A., McBranch, D. W., Leatherdale, C. A. & Bawendi, M. G. Quantization of multiparticle Auger rates in semiconductor quantum dots. *Science* **287,** 1011-1013 (2000).
30. Lenngren, N. *et al.* Multiexciton absorption cross sections of CdSe quantum dots determined by ultrafast spectroscopy. *J. Phys. Chem. Lett.* **4,** 3330-3336 (2013).
31. Ji, M. *et al.* Efficient multiple exciton generation observed in colloidal PbSe quantum dots with temporally and spectrally resolved intraband excitation. *Nano Lett.* **9,** 1217-1222 (2009).
32. Amarotti, E. *et al.* Direct visualization of confinement and many-body correlation effects in 2D spectroscopy of quantum dots. *Adv. Opt. Mater.* **12,** 2302968 (2024).
33. Kambhampati, P. Unraveling the structure and dynamics of excitons in semiconductor quantum dots. *Acc. Chem. Res.* **44,** 1-13 (2011).

34. Klimov, V. I. Spectral and dynamical properties of multiexcitons in semiconductor nanocrystals. *Annu. Rev. Phys. Chem.* **58,** 635-673 (2007).
35. Nozik, A. J. Multiple exciton generation in semiconductor quantum dots. *Chem. Phys. Lett.* **457,** 3-11 (2008).
36. Chen, J., Messing, M. E., Zheng, K. & Pullerits, T. Cation-dependent hot carrier cooling in halide perovskite nanocrystals. *J. Am. Chem. Soc.* **141,** 3532-3540 (2019).
37. Ben-Shahar, Y. *et al.* Charge carrier dynamics in photocatalytic hybrid semiconductor–metal nanorods: Crossover from Auger recombination to charge transfer. *Nano Lett.* **18,** 5211-5216 (2018).
38. Tahara, H., Sakamoto, M., Teranishi, T. & Kanemitsu, Y. Quantum coherence of multiple excitons governs absorption cross-sections of PbS/CdS core/shell nanocrystals. *Nat. Commun.* **9,** 3179 (2018).
39. Puthenpurayil, J., Cheng, O. H.-C., Qiao, T., Rossi, D. & Son, D. H. On the determination of absorption cross section of colloidal lead halide perovskite quantum dots. *J. Chem. Phys.* **151,** (2019).
40. Zheng, K. *et al.* High excitation intensity opens a new trapping channel in organic–inorganic hybrid perovskite nanoparticles. *ACS Energy Lett.* **1,** 1154-1161 (2016).
41. Lüttgens, J. M., Kuang, Z., Zorn, N. F., Buckup, T. & Zaumseil, J. Evidence for a polariton-mediated biexciton transition in single-walled carbon nanotubes. *ACS Photonics* **9,** 1567-1576 (2022).
42. McKillop, A. M., Chen, L., Fidler, A. P. & Weichman, M. L. Direct readout of excited-state lifetimes in chlorin chromophores under electronic strong coupling. *J. Am. Chem. Soc.* **148,** 9737-9753 (2026).
43. Wang, S. *et al.* Quantum yield of polariton emission from hybrid light-matter states. *J. Phys. Chem. Lett.* **5,** 1433-1439 (2014).
44. Ahn, W., Vurgaftman, I., Dunkelberger, A. D., Owrutsky, J. C. & Simpkins, B. S. Vibrational strong coupling controlled by spatial distribution of molecules within the optical cavity. *ACS Photonics* **5,** 158-166 (2018).
45. Schouwink, P., Berlepsch, H. v., Dähne, L. & Mahrt, R. F. Dependence of Rabi splitting on the spatial position of the optically active layer in organic microcavities in the strong coupling regime. *Chem. Phys.* **285,** 113-120 (2002).
46. Byrnes, S. J. Multilayer optical calculations. arXiv:1603.02720 (2016).

47. Gomes Ferreira, M. *et al.* Self-assembly of quantum-confined $CsPbBr_3$ perovskite nanocrystals into rhombic, frame, and rectangular superlattices. *Small Structures* **6,** 2500133 (2025).
48. Dong, Y. *et al.* Precise control of quantum confinement in cesium lead halide perovskite quantum dots via thermodynamic equilibrium. *Nano Lett.* **18,** 3716-3722 (2018).